# Direct Probing of the Stacking Order and Electronic Spectrum of Rhombohedral Trilayer Graphene with Scanning Tunneling Microscopy

Rui Xu[§], Long-Jing Yin[§], Jia-Bin Qiao[§], Ke-Ke Bai, Jia-Cai Nie, Lin He*

Department of Physics, Beijing Normal University, Beijing, 100875, People's Republic of China

**Recently, the rhombohedral trilayer graphene (r-TLG) has attracted much attention because of its low-energy flat bands, which are predicted to result in many strongly correlated phenomena. However, experimental evidence for these flat bands in the r-TLG is still missing so far since that the supporting substrates usually have strong destructive effects on the low-energy band structure of graphene systems. Here, we demonstrate that it is possible to direct probe the stacking order and electronic spectrum of the r-TLG on a graphite surface with a scanning tunneling microscopy around a monoatomic step edge of the top graphene layer. The tunneling spectra of the r-TLG exhibit four adjacent peaks, which are generated by the low-energy flat bands, flanking the charge neutrality point. Based on these spectra, the true energy gap and the energy gap at the K-point of the r-TLG are determined as about 9 meV and 23 meV, respectively. The observed features are well reproduced by a low-energy effective Hamiltonian.**

[§]These authors contributed equally to this work.

*Email: helin@bnu.edu.cn

Scanning tunneling microscopy (STM) is a powerful technique that has achieved great success in establishing close connections between crystal structures and electronic properties of graphene monolayer and bilayer [1-6]. In graphene multilayers, such as graphene bilayer and trilayer, the stacking order can dramatically impact their electronic spectra, and recently, it is treated as an unique degree of freedom to tune the electronic properties of graphene systems [3,7-22]. As an example, placing a graphene monolayer on top of a Bernal graphene bilayer will form two natural stable allotropes of the trilayer: one is the Bernal-stacked (or ABA) trilayer, the other is the rhombohedral-stacked trilayers (r-TLG or ABC trilayer). The subtle difference in stacking order of the two systems results in dramatic distinctions in their band structures and electronic properties [7-14]. Very recently, the r-TLG has attracted much attention because of its low-energy flat bands, which are predicted to result in many strongly correlated phenomena [7,23-26], such as spontaneous gap formation, superconductivity, and ferromagnetism. However, the STM fails to identify the stacking order for the graphene multilayers with the layer number $N \geq 3$. To distinguish the two natural stable allotropes of the trilayer graphene, it was believed that we have to seek help from other techniques [7-14]. Additionally, a more serious obstacle for accessing the electronic structure of the trilayer graphene is the fact that the substrates usually have strong destructive effects on the low-energy band structures of the graphene systems [7-14,27].

In this Letter, we demonstrate that it is possible to probe the stacking order of the r-TLG on a graphite surface directly with a STM around a monoatomic step edge of the top graphene layer. As demonstrated previously, flat bands are manifested in pronounced peaks in tunneling density of states (DOS) [1,6,28-30]. Thus we also seek to experimentally probe the low-energy electronic structure of the r-TLG with the STM. In the tunneling spectra of the r-TLG, four peaks flanking the charge neutrality point, which are generated by the low-energy flat bands [15-17], are observed. The true energy gap and the energy gap at the K-point of the r-TLG [16] are determined as about 9 meV and 23 meV, respectively.

Our experiments were performed on highly oriented pyrolytic graphite (HOPG) surface. The HOPG sample was of ZYA grade (NT-MDT) and it was surface cleaved with adhesive tape prior to experiments. There are two main advantages of using HOPG. First, rhombohedral graphite accounts for a few part of HOPG [31] and the surface few layers decouple from the bulk HOPG sample [3,32-35]. Therefore, the surface of HOPG provides a natural ideal platform to probe the electronic spectra of the r-TLG. Second, according to our result presented here, graphite is a promising substrate in maintaining the low-energy band structure of graphene systems. This provides unprecedented opportunities for accessing the low-energy fine spectra of the r-TLG. The STM system was an ultrahigh vacuum scanning probe microscope (USM-1500) from UNISOKU (See Supplemental Material [36] for experimental method).

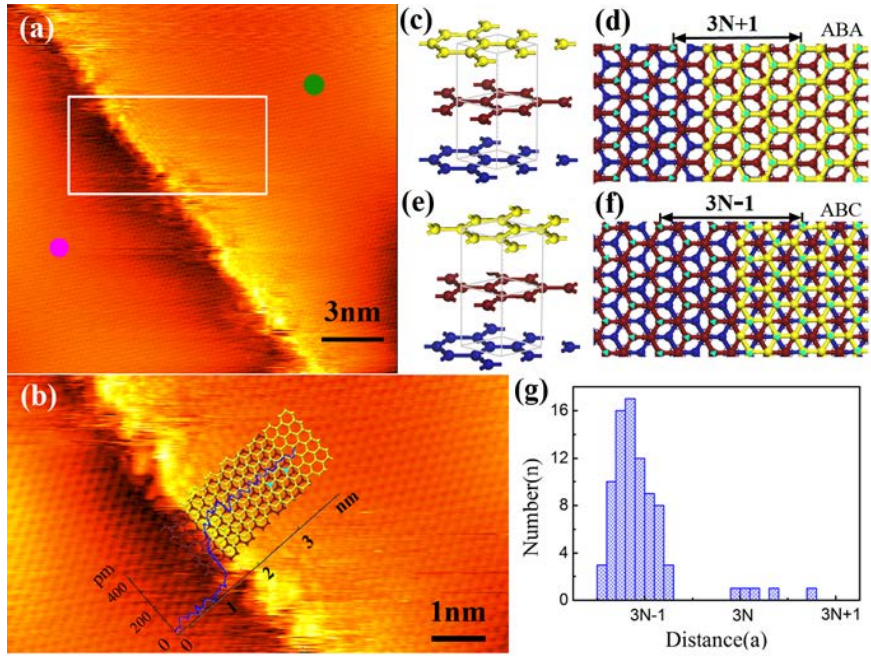

**Figure 1** (color online). STM images of a r-TLG around a monoatomic step edge and its schematic structure. (a) A STM image of HOPG surface around a monoatomic step with the zigzag edge ($V_{sample}$ = 0.80 *V*, I = 0.129 *nA*). (b) Zoom-in topography of the white frame in panel (a) ($V_{sample}$ = 0.90 *V*, I = 0.129 *nA*). The blue curve is a typical section line across the step. The atomic structures of the first and second graphene layers around the step edge are overlaid onto the STM image. The topmost layer shifts by an atomic spacing with respect to the second layer along the perpendicular direction of the step edge. (c) (e) Schematic side views of ABA-stacked and ABC-stacked TLG, respectively. (d) (f) Schematic top views of ABA-stacked and ABC-stacked TLG, respectively. The cyan triangles represent the bright dots of triangular lattice in the STM images. In ABA-stacked [ABC-stacked] TLG, the distance between any two bright dots-which are flanking the step edge and on a line along the perpendicular direction of the edge-is (3N+1)a [(3N−1)a] with a = 0.142 nm and N a positive integer. (g) The measured distribution of the distance between the two bright dots, as defined in (d) and (f), in the experiment.

Figure 1(a) shows a STM image obtained on the surface of a HOPG. There is a monoatomic step edge on the surface and a small portion of the zigzag edge is almost crystallographically perfect, as shown in Fig. 1(b). Both the left and the right regions of the edge show a triangular contrast in the STM images because of the A/B atoms asymmetry generated by the adjacent two layers. These bright spots of the triangular lattice correspond to the sites in the adjacent two layers where one sublattice of the first layer lies above the center of the hexagons in the second layer. Usually, the main part of HOPG is the Bernal graphite and the rhombohedral graphite only accounts for a few part of the sample. According to STM images, it is difficult to distinguish the rhombohedral graphite and the Bernal graphite since that both of them show the triangular lattice in the images. Thanks to the crystallographically perfect edge, as shown in Fig. 1(b), we can determine the exact atomic arrangement of the first and the second graphene layers, as shown in the inset of Fig. 1(b). Then, for the topmost three layers in the right region of the edge, we could differentiate the r-TLG from the Bernal-stacked TLG by carefully measuring the triangular lattice around the zigzag step edge, i.e., we can identify the relative shift of the third layer with respect to the second layer quantitatively, as schematic shown in Fig. 1(c)-1(f). In the r-TLG, the distance between any two bright dots of the triangular lattice, which are flanking the edge and on a line along the perpendicular direction of the edge, should be $(3N-1)a$ with $a = 0.142$ nm and $N$ a positive integer. In the Bernal-stacked TLG, it should be $(3N+1)a$ (See Supplemental Material [36] for a typical result of the Bernal-stacked TLG on the graphite surface). Such a subtle distinction is large enough in the STM measurement because of its ultra-high spatial resolution. Figure 1(g) shows histogram of the measured data, which exhibit a distribution of values around (3$N$-1)a. Therefore, we can conclude that the topmost three layers in the right region of the zigzag edge is the r-TLG.

To further ascertain that the topmost three layers in the right region of the edge correspond to the r-TLG, we carried out STS measurements. Figure 2(a) and 2(b) show two typical spectra measured in the left region and right region of the zigzag edge, respectively. To remove any possible effect of the zigzag edge [28,29,37,38], the spectrum in the right region is obtained at about 8 nm away from it. While the spectrum in the left region shows no discernible structure, low-energy pronounced peaks are clearly resolved in the tunneling spectrum measured in the right region. The tunneling spectrum gives direct access to the local DOS of the surface. Therefore, these low-energy tunneling peaks are attributed to the signatures of the flat bands around the charge neutrality point of the r-TLG [15-17].

The inset of Fig. 2(b) shows a high resolution (2 mV) spectroscopy around the charge neutrality point of the r-TLG. To ascertain the reproducibility of the results, several tens spectra of the r-TLG are recorded. At a fixed position, the basic features of the tunneling spectra are quite reproducible. At different positions, the peak heights and peak positions of the spectra vary very slightly, the origin of which will be discussed subsequently. The substrate (the bulk HOPG here) could induce an effective external electric field on the r-TLG [8], which generates a finite band gap in the r-TLG [8,16], as shown in Fig. 2(c) and

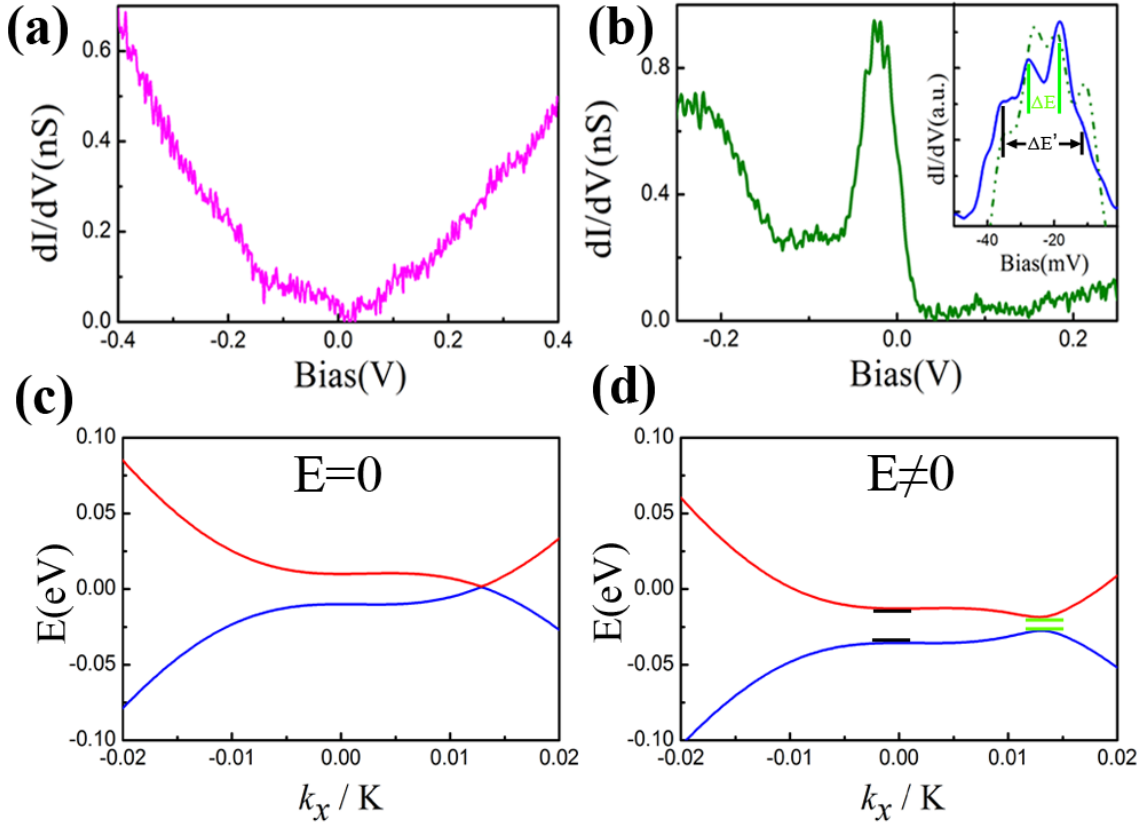

**Figure 2** (color online). STS spectra and electronic structures of the r-TLG. (a) The STS taken at the magenta dot in Fig. 1(a). (b) The STS taken at the green dot in Fig. 1(a). A zoom-in of the tunneling peaks in the spectrum (dark green dash-dotted curve) is shown in the inset. The inset also shows a high-resolution STS (blue curve), which exhibits four tunnelling peaks around the charge neutrality point of the r-TLG. All the spectra are measured at about 4.5 K. (c) Band structure for r-TLG with zero electric field. (d) Band structure for r-TLG with a moderately sized electric field. The true energy gap ΔE and the energy gap at the K-point ΔE' of the r-TLG are indicated by green and black lines, respectively. The low-energy band structure direct results in the fine structures of the tunneling peaks obtained in (b).

2(d) (see Supplementary Material [36] for details of calculation). Two peaks in the spectrum with energy difference of about 9 meV are signatures of the van Hove singularities at the conduction band and valence band edges, as shown in Fig. 2(b) and 2(d). The charge neutrality point $N_C$, which locates at the center of the two peaks, of the r-TLG is determined as about -25 meV. The other two peaks in the spectrum with energy difference of about 23 meV are features of the band gap at the K-point $\Delta E'$, where the energy bands are rather flat. The theoretical result calculated by the low energy Hamiltonian [15-17] agrees with our experimental data quantitatively.

The spectra in Fig. 2(a) and Fig. 2(b) also indicate that the chemical potentials in the left region and right region of the zigzag edge differ strongly. Such a result is attributed to spatial charge inhomogeneity of the sample. Similar spatial charge inhomogeneity, which can be caused by electron-hole puddle, topographic corrugations, charge-donating impurities, and substrates, has been observed previously in many different graphene systems [39-42]. In our experiment, the difference of the microscopic chemical potentials between the left region and the right region may mainly arise from the charge donation of the zigzag edge in the top layer of the r-TLG.

We now turn to investigate effects of the zigzag edge on the electronic properties of the r-TLG. Figure 3(a) shows an enlarged STM image around the zigzag edge and Fig. 3(b) shows position-dependent spectra recorded along a line in the perpendicular direction of the edge. The graphene zigzag edge has localized states located at the conduction band and

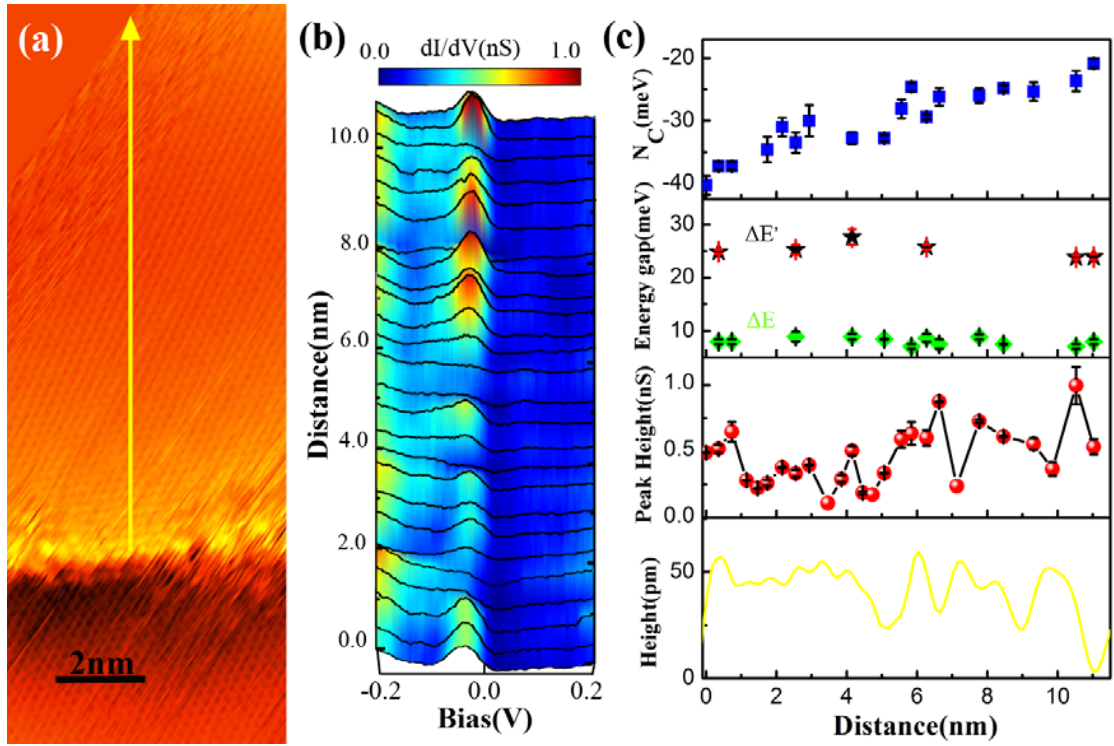

**Figure 3** (color online). Position-dependent STS spectra of the r-TLG. (a) An enlarged STM image of the r-TLG around the zigzag edge ($V_{sample}$ = 0.80 V, $I$ = 0.129 nA). (b) Tunneling spectra acquired at different positions along the yellow arrow in (a). (c) From top to bottom: the charge neutrality points, $N_C$, determined from the tunneling curves in (b) versus the distance from the edge; the true energy gap ΔE and the energy gap at the K-point ΔE' of the r-TLG as a function of the distance from the edge; the height of the low-energy maximum peak of each tunneling curve in (b) versus the distance from the edge; a section line along the yellow arrow in (a).

valence band edges [28,29,37,38], which are expected to result in two low-energy peaks of the local DOS. The electronic density of the edge states is localized at the edge [28,29,37,38] and almost decreases to zero at about 4-6 nm away from the edge. Therefore, around the zigzag edge, both the localized states of the edge and the van Hove singularities of the r-TLG should contribute to the tunneling peaks around the charge neutrality point. Because of the superposition of their located energies, it is difficult to separate their contributions in the tunneling spectra measured near the edge (see Supplementary Material [36] for enlarged image of the experimental data). However, when the distance away from the edge is larger than 6 nm, it is reasonable to neglect the contributions from the edge states.

Figure 3(c) summarizes several physical parameters, which are derived from the tunneling spectra and the STM image, as a function of the distance from the edge. According to the position-dependent charge neutrality point, the zigzag edge generates a local microscopic charge inhomogeneity in the r-TLG around the edge. Such a result is reasonable because of that the sublattice symmetry of graphene is broken at the edge and many adatoms (impurities) may hybridize directly with carbon atoms at the edge. The charge-donating zigzag edge not only leads to the the position-dependent charge neutrality point in the r-TLG, but also account for the difference of the chemical potentials between its left region and right region, as shown in Fig. 2(a) and 2(b). The true energy gap and the energy gap at the K-point of the r-TLG, as shown in Fig. 3(c), seem not being affected by the zigzag edge. Both of them are almost independent of the positions (the slight variations of $\Delta E$ and $\Delta E'$ are within the energy resolution of the tunneling spectra). However, the height of the low-energy maximum peak of each tunneling curve depends sensitively on the measured positions, as shown in Fig. 3(c). We attributed this behavior mainly to intrinsic/extrinsic ripples of the topmost graphene layer [43,44]. The bottom panel of Fig. 3(c) shows a typical profile line of the surface graphene exhibiting ripples with wavelength of 1-3 nm and amplitude of 10-40 pm. These ripples could enlarge or reduce the interlayer distance, which consequently could weaken or enhance the interlayer coupling between the first and second graphene layers. This subtle distinction of the interlayer coupling can change the range of $k_x$ values, where the low-energy flat bands are well defined (see Supplementary Material [36] for details of calculation). Therefore, it can affect the intensity of the low-energy tunneling peaks and result in the difference of the spectra at different positions, as observed in our experiment.

In summary, the stacking order and low-energy electronic spectrum of the r-TLG were studied carefully by using STM and STS. The low-energy flat bands of the r-TLG are manifested in the four adjacent peaks of the tunneling spectra. Because the low energy van Hove singularities or the flat bands are vital to many strongly correlated phenomena, therefore, our result indicates that the r-TLG is an ideal platform in exploring novel correlated electronic phases in graphene.

**Acknowledgements**

This work was supported by the National Basic Research Program of China (Grants Nos. 2014CB920903, 2013CBA01603, 2013CB921701), the National Natural Science Foundation of China (Grant Nos. 11422430, 11374035, 51172029, 91121012), the program for New Century Excellent Talents in University of the Ministry of Education of China (Grant No. NCET-13-0054), Beijing Higher Education Young Elite Teacher Project (Grant No. YETP0238), and the Fundamental Research Funds for the Central Universities.

**References**

[1] N. Levy, S. A. Burke, K. L. Meaker, M. Panlasigui, A. Zettl, F. Guinea, A. H. Castro Neto, and A. F. Crommie, *Science* **329**, 544 (2010).

[2] L. Zhao, R. He, K. T. Rim, T. Schiros, K. S. Kim, H. Zhou, C. Gutierrez, S. P. Chockalingam, C. J. Arguello, L. Palova, D. Nordlund, M. S. Hybertsen, D. R. Reichman, T. F. Heinz, P. Kim, A. Pinczuk, G. W. Flynn, and A. N. Pasupathy, *Science* **333**, 999 (2011).

[3] G. Li, A. Luican, J. M. B. Lopes dos Santos, A. H. Castro Neto, A. Reina, J. Kong, and E. Y. Andrei, *Nature Phys.* **6**, 109 (2010).

[4] M. Yankowitz, J. Xue, D. Cormode, J. D. Sanchez-Yamagishi, K. Watanabe, T. Taniguchi, P. Jarillo-Herrero, P. Jacquod, and B. J. LeRoy, *Nature Phys.* **8**, 382 (2012).

[5] L. Tapaszto, T. Dumitrica, S. J. Kim, P. Nemes-Incze, C. Hwang, and L. P. Biro, *Nature Phys.* **8**, 739 (2012).

[6] W. Yan, W.-Y. He, Z.-D. Chu, M. Liu, L. Meng, R.-F. Dou, Y. Zhang, Z. Liu, J.-C. Nie, and L. He, *Nature Commun.* **4**, 2159 (2013).

[7] W. Bao, L. Jing, J. Velasco Jr, Y. Lee, G. Liu, D. Tran, B. Standley, M. Aykol, S. B. Cronin, D. Smirnov, M. Koshino, E. McCann, M. Bockrath, and C. N. Lau, *Nature Phys.* **7**, 948 (2011).

[8] M. Yankowitz, F. Wang, C. N. Lau, and B. J. LeRoy, *Phys. Rev. B* **87**, 165102 (2013).

[9] M. Yankowitz, J. I-J.Wang, A. G. Birdwell, Y.-A. Chen, K. Watanabe, T. Taniguchi, P. Jacquod, P. San-Jose, P. Jarillo-Herrero, and B. J. LeRoy, *Nature Mater.* **13**, 786 (2014).

[10] T. Ohta, A. Bostwick, J. L. McChesney, T. Seyller, K. Horn, and E. Rotenberg, *Phys. Rev. Lett.* **98**, 206802 (2007).

[11] K. F. Mak, J. Shan, and T. F. Heinz, *Phys. Rev. Lett.* **104**, 176404 (2010).

[12] Z. Li, C. H. Lui, E. Cappelluti, L. Benfatto, K. F. Mak, G. L. Carr, J. Shan, T. F. Heinz, *Phys. Rev. Lett.* **108**, 156801 (2012).

[13] C. H. Lui, E. Cappelluti, Z. Li, and T. F. Heinz, *Phys. Rev. Lett.* **110**, 185504 (2013).

[14] L. Zhang, Y. Zhang, J. Camacho, M. Khodas, and I. Zaliznyak, *Nature Phys.* **7**, 953 (2011).

[15] F. Zhang, B. Sahu, H. Min, and A. H. MacDonald, *Phys. Rev. B* **82**, 035409 (2010).

[16] A. A. Avetisyan, B. Partoens, and F. M. Peeters, *Phys. Rev. B* **81**, 115432 (2010).

[17] A. Kumar, and R. Nandkishore, *Phys. Rev. B* **87**, 241108(R) (2013).

[18] E. V. Castro, K. S. Novoselov, S. V. Morozov, N. M. R. Peres, J. M. B. Lopes dos Santos, J. Nilsson, F. Guinea, A. K. Geim, and A. H. Castro Neto, *Phys. Rev. Lett.* **99**, 216802 (2007).

[19] J. M. B. Lopes dos Santos, N. M. R. Peres, and A. H. Castro Neto, *Phys. Rev. Lett.* **99**, 256802 (2007).

[20] W. Yan, M. Liu, R.-F. Dou, L. Meng, L. Feng, Z.-D. Chu, Y. Zhang, Z. Liu, J.-C. Nie, and L. He, *Phys. Rev. Lett.* **109**, 126801 (2012).

[21] W.-Y. He, Z.-D. Chu, and L. He, *Phys. Rev. Lett.* **111**, 066803 (2013).

[22] F. Guinea, A. H. Castro Neto, and N. M. R. Peres, *Phys. Rev. B* **73**, 245426 (2006).

[23] N. B. Kopnin, T. T. Heikkila, and G. E. Volovik, *Phys. Rev. B* **83**, 220503(R) (2011).

[24] H. Liu, H. Jiang, X. C. Xie, and Q.-F. Sun, *Phys. Rev. B* **86**, 085441 (2012).

[25] R. Olsen, R. van Gelderen, and C. Morais Smith, *Phys. Rev. B* **87**, 115414 (2013).

[26] H. Wang, J.-H. Gao, and F.-C. Zhang, *Phys. Rev. B* **87**, 155116 (2013).

[27] S. Hattendorf, A. Georgi, M. Liebmann, M. Morgenstern, *Surf. Sci.* **610**, 53 (2013).

[28] Y. Kobayashi, K.-I. Fukui, T. Enoki, K. Kusakabe, and Y. Kaburagi, *Phys. Rev. B* **71**, 193406 (2005).

[29] Y. Niimi, T. Matsui, H. Kambara, K. Tagami, M. Tsukada, and H. Fukuyama, *Phys. Rev. B* **73**, 085421 (2006).

[30] L. Feng, X. Lin, L. Meng, J.-C. Nie, J. Ni, and L. He, *Appl. Phys. Lett.* **101**, 113113

(2012).

[31] J. Hass, W. A. de Heer, and E. H. Conrad, *J. Phys.: Condens. Matter* **20**, 323202 (2008).

[32] T. Matsui, H. Kambara, Y. Niimi, K. Tagami, M. Tsukada, and H. Fukuyama, *Phys. Rev. Lett.* **94**, 226403 (2005).

[33] G. Li and E. Y. Andrei, *Nature Phys.* **3**, 623 (2007).

[34] D. L. Miller, K. D. Kubista, G. M. Rutter, M. Ruan, W. A. de Heer, P. N. First, and J. A. Stroscio, *Science* **324**, 924 (2009).

[35] L.-J. Yin, J.-B. Qiao, W.-X. Wang, Z.-D. Chu, K. F. Zhang, R.-F. Dou, C. L. Gao, J.-F. Jia, J.-C. Nie, and L. He, *Phys. Rev. B* **89**, 205410 (2014).

[36] See Supplemental Material for more experimental data, details of the calculation, and analysis.

[37] C. Tao, L. Jiao, O. V. Yazyev, Y.-C. Chen, J. Feng, X. Zhang, R. B. Capaz, J. M. Tour, A. Zettl, S. G. Louie, H. Dai, M. F. Crommie, *Nature Phys.* **7**, 616 (2011).

[38] J. Baringhaus, F. Edler, C. Tegenkamp, *J. Phys.: Condens. Matter* **25**, 392001 (2013).

[39] J. Martin, N. Akerman, G. Ulbricht, T. Lohmann, J. H. Smet, K. von K.itzing, A. Yacoby, *Nature Phys.* **4**, 144 (2008).

[40] Y. Zhang, V. W. Brar, C. Girit, A. Zettl, M. F. Crommie, *Nature Phys.* **5**, 722 (2009).

[41] G. M. Rutter, S. Jung, N. N. Klimov, D. B. Newell, N. B. Zhitenev, J. A. Stroscio, *Nature Phys.* **7**, 649 (2011).

[42] R. Decker, Y. Wang, V. W. Brar, W. Regan, H.-Z. Tsai, Q. Wu, W. Gannett, A. Zettl, M. F. Crommie, *Nano Lett.* **11**, 2291 (2011).

[43] J. C. Meyer, A. K. Geim, M. I. Katsnelson, K. S. Novoselov, T. J. Booth, and S. Roth, *Nature* **446**, 60 (2007).

[44] A. Fasolino, J. H. Los, and M. I. Katsnelson, *Nature Mater.* **6**, 858 (2007).